\documentclass[a4paper,twocolumn,%tightenlines,
english,aps,pre,floatfix,groupedaddress,showpacs,nofootinbib]{revtex4-1}
\usepackage[T1]{fontenc}
\usepackage[latin1]{inputenc}
\usepackage{amsmath}
\usepackage{babel}
\usepackage{graphics}
\usepackage{amssymb}
\usepackage{mathrsfs}
\usepackage{dcolumn}
\usepackage{epstopdf}


\begin{document}
\title
{Dynamical aspects of spontaneous symmetry breaking in driven flow with exclusion}
\author {S. L. A. \surname{de Queiroz}}
%\email{sldq@if.ufrj.br}
\affiliation{Instituto de F\'\i sica, Universidade Federal do
Rio de Janeiro, Caixa Postal 68528, 21941-972
Rio de Janeiro, Rio de Janeiro, Brazil}
\author {R. B. \surname{Stinchcombe}}
%\email{Robin.Stinchcombe@physics.ox.ac.uk}
\affiliation{Rudolf Peierls Centre for Theoretical Physics, Clarendon Laboratory,
University of Oxford, Oxford OX1 3PU, United Kingdom}

\date{\today}

\begin{abstract} 
We present a numerical study of a two-lane version of the stochastic 
non-equilibrium model known as the totally asymmetric simple exclusion process. 
For such a system with open boundaries, and suitably chosen values of externally-imposed particle 
injection ($\alpha$) and ejection ($\beta$) rates, spontaneous
symmetry breaking can occur. We investigate the statistics and internal structure
of the stochastically-induced transitions, or "flips", which occur between opposite broken-symmetry 
states as the system evolves in time. From the  distribution of time intervals
separating successive flips, we show that the evolution of the associated characteristic times 
against externally-imposed rates yields information regarding the proximity to a critical
point in parameter space. On short time scales, we probe for the possible existence of
precursor events to a flip between opposite broken-symmetry states. We study an 
adaptation of domain-wall theory to mimic the density reversal process associated with a flip. 
\end{abstract}

%\pacs{05.40.-a, 02.50.-r, 72.80.Vp, 73.23.-b}
%05.40.-a Fluctuation phenomena, random processes, noise, and Brownian motion
%02.50.-r Probability theory, stochastic processes, and statistics
\maketitle
%\tightenlines
 
\section{Introduction} 
\label{intro} 

%%%%%%%%%%%%%%%%% 2nd referee, point (1) [1/2] %%%%%%%%%%%%%%%%%%%%%
Spontaneous symmetry breaking (SSB) is present in many systems
studied both experimentally and theoretically. Some of the most
relevant examples
are magnetization reversal~\cite{bbvb03} and its close
relative, domain nucleation~\cite{bbvb05} in two-dimensional
Ising magnets, and (for fluids) flow reorientation and reversal in 
Rayleigh-B\'enard convection~\cite{sbn02,benzi,jfm}.
%%%%%%%%%%%%%%%%%%%%%%%%%%%%%%%%%%%%%%%%%%%%%%%%%%%%%%%%%%%%%%%%%%%

For one-dimensional systems with finite-range 
interactions in thermal equilibrium at non-zero temperature $T$, 
it is well known that SSB is usually prevented by 
fluctuations. 
%%%%%%%%%%%%%%%%%%% 2nd referee, point (4)  (1/2) %%%%%%%%%%%%%%%%%%%%%
This may easily be seen for cases with discrete symmetry, like the Ising model. 
Here the "modes"  (differences of configurations possible at $T>0$  
from those at $T=0$)  are simple (domain walls) with system-size dependent energy
gap~\cite{peierls}. 
%%%%%%%%%%%%%%%%%%%%%%%%%%%%%%%%%%%%%%%%%%%%%%%%%%%%%%%%%%%%%%%%%
In the more subtle cases with 
continuous symmetry the lower critical dimension for $T \neq 0$ phase 
transitions is raised by the soft character of modes, according to the 
Goldstone argument~\cite{goldstone}. These arguments do not apply to 
non-equilibrium systems. In particular it has been predicted by mean-field 
theory, and verified by numerical 
simulations~\cite{efgm95,pem08,arhs10,sg17,vsg18}, 
that stochastic models  in the  family of the  one-dimensional totally 
asymmetric simple exclusion process  
(TASEP)~\cite{derr98,sch00,mukamel,derr93,rbs01,be07,cmz11} do 
exhibit SSB in especially-designed  implementations.

The one-dimensional TASEP exhibits many non-trivial properties because of its 
collective character. It has been used, often with adaptations, to model a broad 
range of non-equilibrium physical phenomena, from the macroscopic level such
as highway traffic~\cite{sz95} to the microscopic, including sequence
alignment in computational biology~\cite{rb02} 
and current shot noise in quantum-dot chains~\cite{kvo10}.

In the time  evolution  of the $d=1$ TASEP,
the particle number $n_\ell$ at lattice site $\ell$ can be $0$ or $1$, 
and the forward hopping of particles is only to an empty adjacent site. 
In addition to the stochastic character provided by random selection of site 
occupation 
update~\cite{rsss98}, the instantaneous current $J_{\ell\,\ell+1}$ 
across the bond from $\ell$ to $\ell+1$ depends also on the stochastic 
attempt rate, or bond (transmissivity) rate, $p_\ell$, associated with it. 
Thus,  
\begin{equation}
J_{\ell\,\ell+1}= \begin{cases}{n_\ell (1-n_{\ell+1})\quad {\rm with\ probability}\ p_\ell}\cr
{0\qquad\qquad\qquad {\rm with\ probability}\ 1-p_\ell\ .}
\end{cases}
\label{eq:jinst}
\end{equation}

We  take systems with open boundary conditions at both ends. In the usual formulation each end 
is associated with an externally imposed, injection  or ejection, attempt rate, respectively 
$\alpha$ and $\beta$~\cite{derr98,derr93}. Considering  a two-lane model~\cite{dh01,css00}  
one allows two types of particles (denoted here by $+$ and $-$)
such that, say, the $+$ particles move only from left to right, and the $-$ ones from right to
left. Then one has two sets of external rates: $\{\alpha^+,\beta^+\}$, and $\{\alpha^-,\beta^-\}$.
In order to enable SSB to arise, one must symmetrize the setup by making
$\alpha^+=\alpha^-$, $\beta^+=\beta^-$. Details of the dynamics are given in Sec.~\ref{sec:tasep-theo}
below. For the moment we recall that the onset of SSB in the present case is inherently 
associated with the stochastic current- and density fluctuations intrinsic to TASEP-like  
phenomena. 

Our main goal here is to discuss dynamical aspects of the connection between fluctuations and SSB.
For earlier work, see especially Refs.~\onlinecite{sch03,wsg05,gsw07}.

 In Sec.~\ref{sec:tasep-theo} we present the additional rules to be used here 
for two-lane TASEP, which
make possible the onset of SSB. In Sec.~\ref{sec:tasep-num} we initially give details of the
procedures used in our numerical simulations; then we analyse the time evolution of the
TASEP model on long time scales, showing that for the values of $(\alpha,\beta)$ considered
the system exhibits an apparent steady state with broken symmetry.
Stochastic 
fluctuations can induce a reversal, or "flip", in composition of the high- and low-density phases,
between $+$ and $-$. On this time scale the "flip" is essentially instantaneous. We study the
statistics of flips and its dependence on the external parameters by various methods.
Next we turn to shorter time scales and investigate the internal structure of the aforementioned flips,
both by direct visualization of simulational data and by probing for the possible existence of 
smaller "precursor" events to a full reversion in the composition of the high- and low-density
phases. We present an adaptation of domain-wall theory to mimic the density reversal
process associated with a flip. In Section~\ref{sec:conc}, we summarize and discuss our results.

\section{TASEP model: theory}
\label{sec:tasep-theo}

In line with Refs.~\onlinecite{efgm95,pem08}, additionally to Eq.~(\ref{eq:jinst}) for single-lane
processes we adopt the following rules for the coexistence (or not) between $+$ and $-$ particles:

(I)\ a $+$ particle on the left and a $-$ particle on the right of a bond can exchange places with
probability $q>0$: $+\ -\ \Rightarrow\ -\ +$.   

(II)\  At the left end, an injection attempt of a $+$ particle can only take place if the first lattice
site is empty of both $+$ and $-$ particles. Likewise for injection attempts of $-$ particles
into the last site at  the right end. 

It has been shown~\cite{efgm95,pem08} that these conditions are sufficient to trigger SSB
provided that the ejection attempt rate $\beta$ is (i) less than $\alpha$ and (ii) suitably low;
numerically, one needs $\beta \lesssim 1/3$~\cite{efgm95}.

\section{TASEP model: numerics}
\label{sec:tasep-num}

\subsection{Introduction}
\label{sec:num-intro}

We consider chains with $N$ sites, $N-1$ internal bonds, plus two so-called injection/ejection bonds
at either extreme: to the left of the first site, where injection of $+$ particles is attempted with rate 
$\alpha$ and ejection of $-$ particles with rate $\beta$; likewise to the right of site $N$,
for ejection of $+$ particles ($\beta$) and injection of $-$ particles ($\alpha$).

For a structure with $N_b=N+1$ bonds, an elementary time step $t_0$ consists of $N_b$ sequential bond update attempts,  each of these according to the following rules: (1) select a bond at random, say, bond $ij$, connecting sites $i$ and $j$; 
(2) if the chosen bond has a site occupied by a $+$ ($-$) particle on its left (right) and (2a) an empty
 site or (2b) a site occupied by a $-$ ($+$) particle on its right (left), then
(3) move the $+$ ($-$) particle across it  with probability (bond rate)  $p_{ij}$ in case (2a), or
exchange $+$ and $-$ particles across the bond with probability $q$, in case (2b).
If an injection/ejection bond is chosen, step (2) is suitably modified to 
account for the particle reservoir (the corresponding bond rate being, respectively,  $\alpha$ or $\beta$); each time such bond is selected,  the current state of occupation of the internal site to 
which it is attached determines whether it will be used for injection or ejection, 
see rule (II) of Sec.~\ref{sec:tasep-theo}. 

Thus, in the course of one time step, some bonds may be selected
more than once for  examination and some may not be examined at all.
This constitutes the {\it random-sequential update} procedure described
in Ref.~\onlinecite{rsss98}, which is the realization of the usual master equation in continuous time.

%%%%%%%%%%%%%%%%%% 2nd referee, point (5) %%%%%%%%%%%%%%%%%%%%%%%%%%%%%%%%%%%%%%%%
The sublattice parallel update  process used in Ref.~\onlinecite{gsw07} is such that
during one elementary step, all bonds are probed once and just once. 
The injection and ejection bonds are treated exactly in the same way as here.
It differs from our  random sequential process in that, as regards internal bonds,
all (say) odd-numbered ones are updated with probability one, and simultaneously 
(hence the "parallel" label attached); then, the even-numbered bonds undergo 
the same simultaneous update process. A moment's reflection shows that 
exactly the same types of elementary moves, and site occupation states,
are allowed both in our method and in theirs. The same can be said regarding the 
forbidden processes and site occupation states. So, on average over many steps, both
processes give statistically equivalent results.
%%%%%%%%%%%%%%%%%%%%%%%%%%%%%%%%%%%%%%%%%%%%%%%%%%%%%%%%%%%%%%%%%%%%%%%%%%%%%%%%%%%%%%%

Here we use all internal bond rates $p_\ell=1$, see Eq.~(\ref{eq:jinst}), as well as $q=1$,
see rule (I) of Sec.~\ref{sec:tasep-theo}.

\subsection{Flip statistics}
\label{sec:num-flip}

We have focused on the region of the $(\alpha,\beta)$ parameter space where coexistence between high- and low-density phases (hd/ld) takes place, as that is where SSB effects are more
intensely exhibited. For ease of comparison with extant results from Ref.~\onlinecite{efgm95}
we start by taking $\alpha=1$, $\beta=0.15$, deep inside the hd/ld phase.

In addition to system-wide currents
of $\pm$ particles  $J_\pm (t)$, we kept track of the position-averaged densities,
\begin{equation}
\rho_\pm (t) \equiv \frac{1}{N}\sum_{i=1}^N n_i^\pm (t)\ ,
\label{eq:rhodef}
\end{equation}
where the $n_i^\pm(t)$ are a straightforward extension of the $n_\ell$ of Eq.~(\ref{eq:jinst})
to systems with two particle species.

We generally started simulations with an empty lattice. After some time $\delta t_{\rm ss}$ an apparent steady state exhibiting SSB is typically attained, in which the current- and density differences, respectively  $\Delta J(t) \equiv J_+(t)-J_-(t)$ and $\Delta \rho (t) \equiv \rho_+(t)-\rho_-(t)$, are non-zero. Then after some much longer interval 
$\delta t_{\rm flip} \gg \delta t_{\rm ss}$ a "flip"  occurs in which $\Delta J$ and $\Delta \rho$
concurrently reverse sign. As the system evolves stochastically, the high- and low-density
coexisting phases alternate in composition between $+$ and $-$ particles, via successive flips.

In fact, even deep within the hd/ld coexistence region a flip takes place over a succession of 
many elementary time intervals $t_0$ as defined in 
%%%%%%%%%%%%%%% 2nd report of 2nd referee %%%%%%%%%%%%%%%%%%%
the second paragraph of Sec.~\ref{sec:num-intro}. 
%%%%%%%%%%%%%%%%%%%%%%%%%%%%%%%%%%%%%%%%%%%%%%%%%%%%%%%%%%%
So, one usually defines a
"renormalized time" $t_r =N_r\,t_0$ and considers the quantities $J_\pm (t_r)$, $\rho_\pm (t_r)$,
which are averages of the instantaneous $ J_\pm (t)$, $\rho_\pm (t)$ over $N_r$
consecutive elementary intervals $t_0$.

Of course, in the thermodynamic limit $N \to \infty$ one expects the time taken for the system to
switch between different symmetry-broken phases to diverge. 
Conclusive numerical evidence shows that the typical time between consecutive flips diverges
exponentially in $N$ for TASEP-like models such as the one studied 
here~\cite{efgm95,gsw07}.
Thus, in what follows we will generally concentrate on fixed, finite $N$ and investigate the
features exhibited in finite-size flip dynamics.

\begin{figure}
{\centering \resizebox*{3.2in}{!}{\includegraphics*{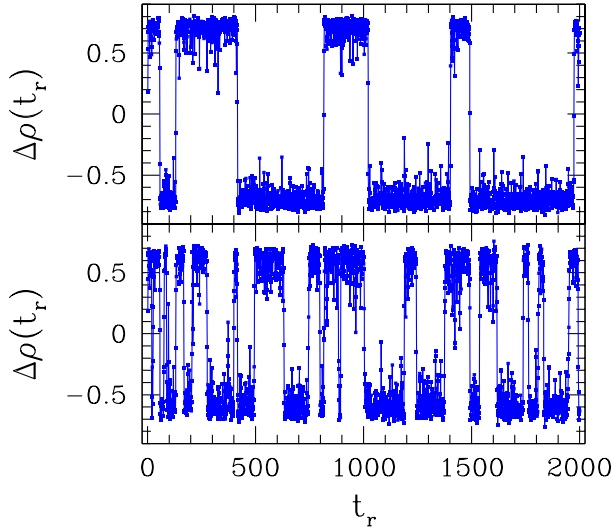}}}
\caption{
Density differences $\Delta\rho(t_r)$ against renormalized time $t_r=10^3\,t_0$ for
chain with $N=80$ sites, $\alpha=1$. Upper panel: $\beta=0.15$; lower panel: $\beta=0.20$.
See text for definitions of $t_r$, $t_0$. 
}
\label{fig:drhoa1b15b20}
\end{figure}

For $(\alpha,\beta)=(1,0.15)$ with system size N=80, using $N_r=10^3$ gives a sharp definition of the 
"instant" (single value of $t_r$) when flips occur, see upper panel of Fig.~\ref{fig:drhoa1b15b20}.
For higher $\beta=0.20$ (but still well away from the transition from hd/ld to the symmetric phase), 
see lower panel of the Figure, comparison with $\beta=0.15$ shows: (i) a reduction of the typical time $\Delta t_r$ between consecutive flips; (ii) the appearance
of  "quasi-flips", in which $\Delta \rho$ goes slightly beyond zero but backtracks before full
reversal; and (iii) the amplitude of flips slightly decreases, as generally expected for the order
parameter of a system approaching a second-order phase transition (in this case,
located at $\beta \approx 1/3$~\cite{efgm95}). 

\begin{figure}
{\centering \resizebox*{3.2in}{!}{\includegraphics*{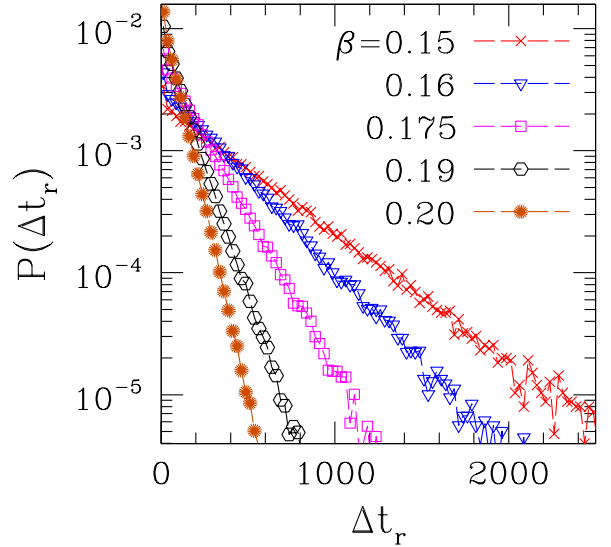}}}
\caption{
For $\alpha=1$ and assorted $\beta$ in the hd/ld SSB phase, log-linear plots
of numerically-evaluated PDFs of renormalized time intervals $\Delta t_r$
between consecutive flips for chain with $N=80$ sites,  $t_r=10^3\,t_0$.
System size $N=80$. See text for definitions of $t_r$,~$t_0$. 
}
\label{fig:pdtr}
\end{figure}

We performed a systematic investigation of point (i), by numerically evaluating the probability
distribution functions (PDF) of the renormalized time intervals $\Delta t_r$
for $\alpha=1$ and $\beta=0.15$, $0.16$, $0.175$, $0.19$, and $0.20$. 
Results are shown in Fig.~\ref{fig:pdtr}.

For all cases investigated the distributions are very well fitted by
a Poissonian form, 
\begin{equation}
P(\Delta t_r) = \frac{1}{\tau_0}\,\exp\left(-\frac{\Delta t_r}{\tau_0}\right)\ .
\label{eq:poisson}
\end{equation} 
Poisson-like distributions for the time between consecutive flips have 
been found in Ref.~\onlinecite{gsw07}, for a slightly different model whose main 
qualitative features are similar to those of the one studied here.

The $\beta-$dependent numerical values of the characteristic times $\tau_0$
are given in Table~\ref{t1}. Note that all fits exclude very short times $\Delta t_r < 20$. This
is because so-called "quasi-flips" defined above, while corresponding to very short-lived sign
changes of $\Delta \rho$ without reaching full (quasi-stable) population reversion, 
tend to contaminate the statistics there  (though they represent  a very small fraction of the total
number of sign-change events, for the interval of $\beta$ under discussion;
see the similar case of "spurious" magnetization reversals of 
two-dimensional Ising spin systems, treated in Ref.~\onlinecite{bbvb03}). 

Our samples
were  of total length $2 \times 10^7$ in units of $t_r$, so the last column in the Table 
shows that the frequency of sign-change events remains rather low, reaching at most $1.5\%$
for $\beta=0.20$.  
 
\begin{table}
\caption{\label{t1} Results of fits of data shown in Fig.~\ref{fig:pdtr} to the Poissonian form,
Eq.~(\ref{eq:poisson}). See text for remarks on fit ranges and definition of numbers of events.
}
\begin{ruledtabular}
\begin{tabular}{@{}cccc}
$\beta$ & $\tau_0$  & Fit range & \# events  \\
\hline\noalign{\smallskip}
$0.15$  &   $419\,(2)$ & $20-2500$ &  $50106$ \\
$0.16$  &    $301\,(2)$ & $20-1900$ &  $71398$ \\
 $0.175$ &   $175\,(1)$ & $20-1500$ &  $123116$ \\
 $0.19$ &   $105\,(2)$ &  $20-900$ &  $211726$ \\
 $0.20$  &   $72\,(2)$  &   $20-750$ &  $308370$ \\
\end{tabular}
\end{ruledtabular}
\end{table}

\begin{figure}
{\centering \resizebox*{3.2in}{!}{\includegraphics*{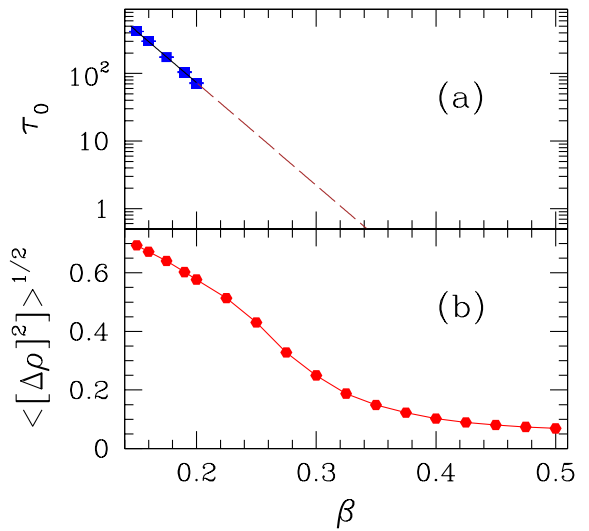}}}
\caption{
(a) On a log-linear plot, the points are
$\tau_0$ from Table~\ref{t1}; the line is the result of fitting data from the Table to
$\tau_0 \propto \exp (-\beta/\delta\beta_0)$, and is extended to $\beta > 0.20$ (dashed section).
(b) Points are RMS values $\langle [\Delta\rho]^2\rangle^{1/2}$ from simulations along 
(renormalized)  time windows of width $t_r=2000$, against $\beta$.
}
\label{fig:drhorms}
\end{figure}

The decay of characteristic times $\tau_0$ upon increasing $\beta$ is fitted very closely by
an exponential form, $\tau_0 \propto \exp(-\beta/\delta\beta_0)$, with $\delta\beta_0 \approx 0.029$.
See part (a) of Fig.~\ref{fig:drhorms}, where the fitted curve is extended to higher $\beta$.
At $\beta=0.20 + 4\delta\beta_0 \approx 0.32$ the predicted value for $\tau_0$
becomes of order one. Although that still corresponds to some $10^3$ elementary updates,
it is very small within the coarse-grained approach used in the analysis which produced the fitted
data. Accordingly, it indicates that such picture of well-defined flips, separated by
relatively long-lasting, quasi-stationary states of phase coexistence, is already breaking
down by $\beta \approx 0.30$ or thereabouts. This is in broad agreement with 
$\beta_c \approx 1/3$ of Ref.~\onlinecite{efgm95}.

In order to reach a better understanding of this behavior, we evaluated the averaged
RMS values $\langle[\Delta \rho]^2\rangle^{1/2}$, where the angular brackets denote averages
over a typical window of width $t_r=2000$, see e.g. Fig.~\ref{fig:drhoa1b15b20}, for
$\alpha=1$, $0.15 \leq \beta \leq 0.50$. Results are shown in part (b) of Fig.~\ref{fig:drhorms}.
It is seen that, as $\beta$ increases and the transition from
hd/ld coexistence to a symmetric phase takes place, the contribution to
 $\langle [\Delta\rho]^2\rangle^{1/2}$ from the diminishing gap between majority- and 
minority phases becomes smaller, and only the nonvanishing, intrinsic stochastic fluctuations
remain. The change of behavior is signalled by the inflection of the curve, located in
the vicinity of $\beta =0.30$, again consistent with the preceding analysis and with
Ref.~\onlinecite{efgm95}.  

We also applied Fourier analysis to our simulation data, in order to quantify the properties
of apparent periodic structures for the signal $\Delta \rho$ which can be seen, e.g., 
in the upper panel of  Fig.~\ref{fig:drhoa1b15b20}. For $\alpha=1$ and $0.15 \leq \beta \leq 0.225$, 
we concentrated  on low frequencies and, for each $\beta$, closely examined the 32 largest-wavelength components  over a total for $M=2048$ sampled points. Our results are more easily conveyed with the help of
the following procedures; (i) with $a_R(f)$, $a_I(f)$ being respectively the real and imaginary
part of the Fourier transform at frequency $f$, we considered the power-like
quantity $a^2_R(f_n)+a^2_I(f_n)$ to represent the contribution of each frequency $f_n=nf_0/M$ 
to the observed signal (here $f_0$ is the highest sampled frequency, equal to one unit of
inverse renormalized time; $n=1,2, \dots, M$); (ii) with the help of standard graphics software, we
smoothed out the resulting sequences of points, in order to average out rapid oscillations and
keep only the basic overall trends.

\begin{figure}
{\centering \resizebox*{3.2in}{!}{\includegraphics*{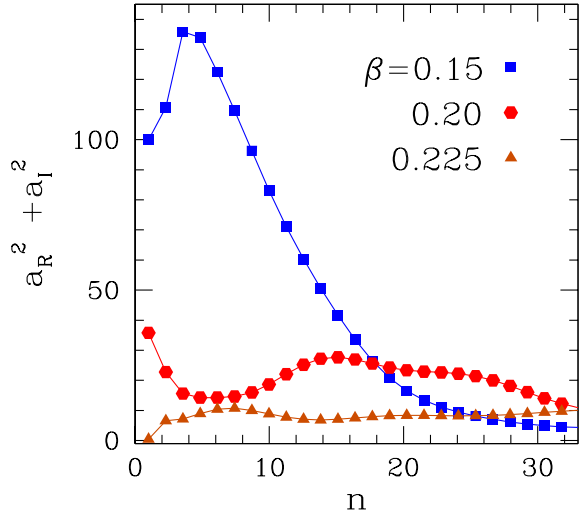}}}
\caption{
For a chain with $N=80$ sites, $\alpha=1$ and varying $\beta$, low-frequency 
results from Fourier analysis of $\Delta \rho (t)$. $a_R$, $a_I$ are, respectively, 
real and imaginary (frequency-dependent) components of Fourier transform. Frequencies 
are denoted by the label $n$  ($f_n=nf_0/M$, with $n=1, 2, \dots, M$; $f_0=1$ in units
of inverse renormalized time, $M=2048$). 
}
\label{fig:fft}
\end{figure}

Results are displayed in Fig.~\ref{fig:fft} for $\beta=0.15$, $0.20$, and $0.225$.
Comparison with the upper panel of Fig.~\ref{fig:drhoa1b15b20} shows that for $\beta=0.15$ the 
$n=3$ and $4$ modes (translating to oscillation periods of order $400-600$ in renormalized time units)
indeed have a strong contribution. For $\beta=0.20$ there is a small but discernible maximum
centered at frequencies around $n=15$, which is roughly in line with data from the 
lower panel of  Fig.~\ref{fig:drhoa1b15b20}. For $\beta=0.225$ no clear trend is visible.

The Poissonian character exhibited by the distributions of time intervals between
flips strongly suggests that these are uncorrelated events, at least on long time scales. We
now proceed to probing the complementary, short-time, scale in order to obtain detailed
information 
on the structure and properties of the fluctuations involved in the overall switching 
between majority-population phases.

\subsection{Flip structure}
\label{sec:num-flip_struc}

\begin{figure}
{\centering \resizebox*{3.2in}{!}{\includegraphics*{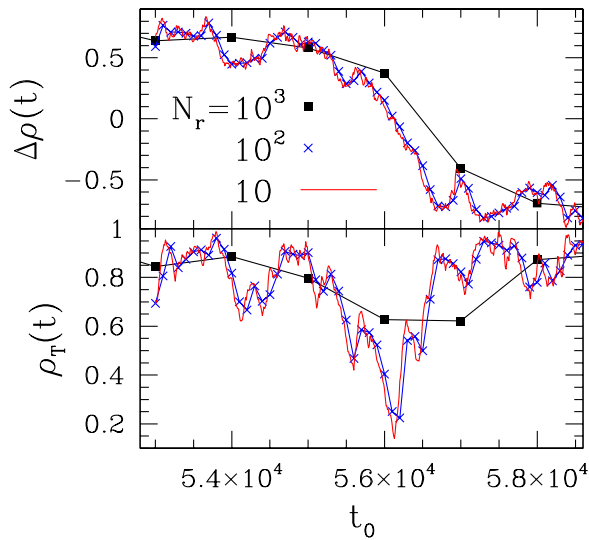}}}
\caption{
Density differences $\Delta\rho(t)$ (upper panel) and total densities $\rho_{\,T}(t)$ (lower panel)
against time $t_0$ in the vicinity of a flip  for chain with $N=80$ sites, $(\alpha,\beta)=(1,0.15)$, 
as seen from averages taken over intervals of lenghts $N_r\,t_0$, $N_r=10^3$, $10^2$ and $10$.  
See text for definition of $t_0$. 
}
\label{fig:a1b15_trvar}
\end{figure}

We begin by examining the features exhibited by a flip when it is seen on different timescales,
i.e., for different values of $t_r= N_r\,t_0$ as defined above. We take $N=80$, $(\alpha,\beta)
=(1,0.15)$. We refer to the leftmost $+ \to -$ flip in the top panel of Fig.~\ref{fig:drhoa1b15b20},
which takes place at $t_r \approx 50-60$. In 
Fig.~\ref{fig:a1b15_trvar} 
we reproduce both $\Delta \rho (t)$ and the total density 
$\rho_{\,T}(t)= \rho_+(t)+\rho_-(t)$ in that time region, for $N_r=10$, $10^2$, and $10^3$.
While the overall trend of $\Delta \rho (t)$ is already well-represented by the $N_r=10^3$
curve, the corresponding data for $\rho_{\,T}(t)$ miss out on a significant 
dip taking place around $t_0=5.6 \times 10^4$, which is captured by the $N_r=10^2$ and $10$ curves. 
Such an almost complete emptying of the lattice is indeed a qualitative feature broadly expected 
to take place during the flipping process~\cite{efgm95,wsg05,gsw07}.

\begin{figure}
{\centering \resizebox*{3.2in}{!}{\includegraphics*{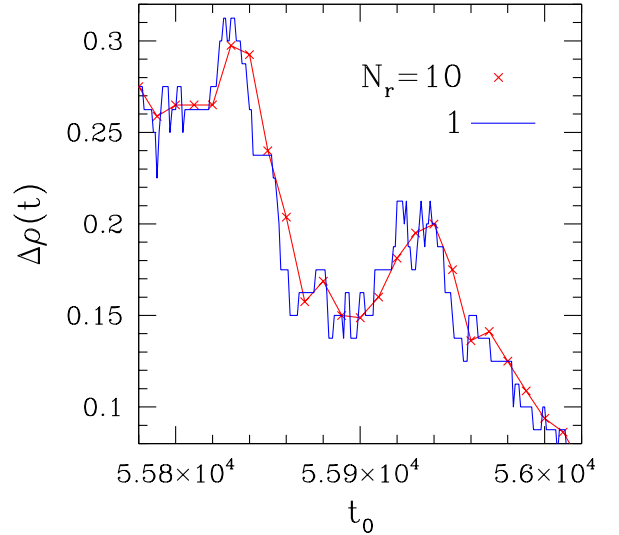}}}
\caption{
Density differences $\Delta\rho(t)$ against time $t_0$ in the vicinity of a flip  for chain with 
$N=80$ sites, $(\alpha,\beta)=(1,0.15)$, 
as seen from averages taken over intervals of length $N_r\,t_0$, $N_r=10$ and $1$.  
See text for definition of $t_0$. 
}
\label{fig:a1b15_trvar2}
\end{figure}

Fig.~\ref{fig:a1b15_trvar2} shows $\Delta\rho(t)$ along a subsection of the time interval
depicted in Fig.~\ref{fig:a1b15_trvar}, now for $N_r=10$ and unity. One sees that, in this case,
no significant insight into the flipping process itself is gained by going to the shortest possible
time scale. On such scale, for the values of $\alpha$, $\beta$ and $N$ used here lattice discreteness 
effects show up to a large extent.
Namely, plateaus are seen corresponding to time 
lapses with no change in overall particle number (this is directly confirmed by analysis of the 
respective $\rho_{\,T}(t)$), as well as step-like variations in density-associated quantities. 
Discreteness effects would be reduced by going to larger $N$ and/or 
higher $\beta$ closer to $\beta_c \approx 1/3$~\cite{efgm95}.

We investigated the possible existence of "precursor" events to flips. These, if at all present, 
would bear a qualitative similarity to the mechanism described in 
Refs.~\onlinecite{wsg05,gsw07} in which, starting from an  empty lattice the
system enters a symmetry-broken state through an "amplification loop" of initial fluctuations.
In the present case we specialize to flips from one symmetry-broken state to its opposite,
for which inertial effects (absent for empty-lattice and other symmetric starting configurations) 
are expected to be paramount throughout the switching process.

\begin{figure}
{\centering 
\resizebox*{3.2in}{!}{\includegraphics*{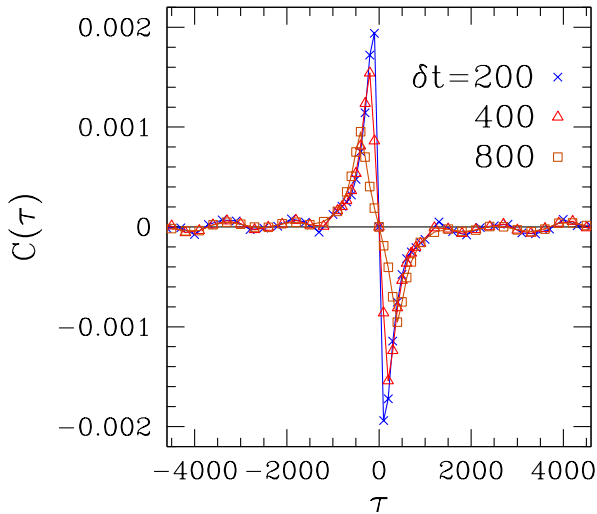}}}
\caption{
Two-time correlation $C(\tau)$ of Eq.~(\ref{eq:ctau_def}).
Densities are averaged over $N_{\rm sam}=10^4$ time intervals of 
length $t_r=10^2 t_0$. 
$\tau$ and $\delta t$ are shown in units of $t_0$. Chain with 
$N=80$ sites, $(\alpha,\beta)=(1,0.15)$. 
See text for definitions of  $N_{\rm sam}$, $t_0$, $\delta t$. 
}
\label{fig:cor_a1b15}
\end{figure}

Examination of the lower panel in Fig.~\ref{fig:a1b15_trvar} suggests that a reliable indicator
for a flip is the steep descent of $\rho_{\,T}(t)$ towards a minimum value much lower
than its long-time average away from flips, $\langle \rho_{\,T} \rangle$. We investigate the
behavior of the two-time correlation (averaged over time $t$): 
\begin{equation}
C(\tau) \equiv \Bigl< \frac{d\rho_{\,T}(t+\tau)}{dt}\,\left( \rho_{\,T}(t)-\langle \rho_{\,T} 
\rangle\right)\Bigr>_t\ .
\label{eq:ctau_def}
\end{equation}   
In order to capture the relevant fluctuations in the neighborhood of a flip, we consider both
positive and negative $\tau$, with $|\tau|$ not much larger than the average "duration" of a 
flip (i.e. the interval around the flip during which $\rho_{\,T}$ differs significantly from its
long-time average); this is estimated from Fig.~\ref{fig:a1b15_trvar} to be a couple of thousand 
times the elementary unit $t_0$. For evaluation of the densities involved in 
Eq.~(\ref{eq:ctau_def}) we used a renormalized time scale $t_r=10^2 t_0$, and
an interval of width $200 \leq\delta t \leq 800$ (in units of $t_0$) 
for numerical calculation of the 
time derivatives. Results from averaging over $N_{\rm sam}=10^4$ samples, 
i.e., consecutive values of $t_r$, are shown 
in Fig.~\ref{fig:cor_a1b15}. Because of the finite values of $\delta 
t$ all curves have peaks at
$\tau=\pm \delta t/2$. Other than the central peaks just mentioned, one sees only a few
small but well-defined secondary oscillations at larger $|\tau|$. The most relevant feature
of the plots for our current purposes is their remarkable degree of inversion symmetry 
around the origin, which strongly indicates that pre- and post-flip fluctuations are,
on average, equivalent. In other words, we have found no specific precursor events 
associated with a flip from one symmetry-broken state to its opposite. 
 
\begin{figure}
{\centering 
\resizebox*{3.2in}{!}{\includegraphics*{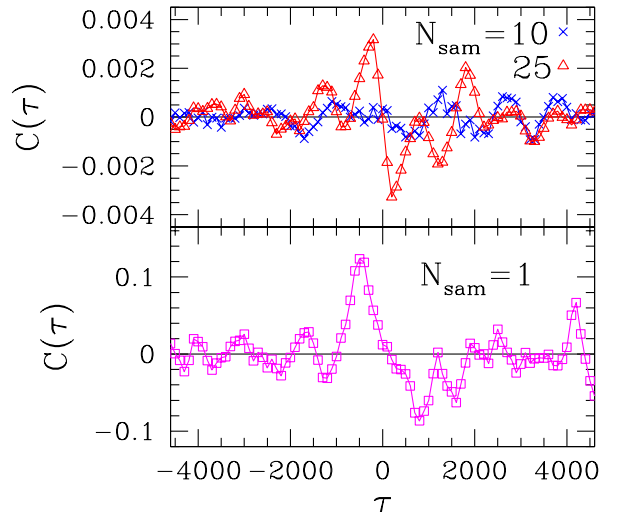}}}
\caption{
Two-time correlation $C(\tau)$ of Eq.~(\ref{eq:ctau_def}).
Densities are averaged over time intervals of length $t_r=10^2 t_0$,
using $\delta t=400$, with $N_{\rm sam}=10$ and $25$ (top panel),
or for a single sample, $N_{\rm sam}=1$ (bottom panel). 
$\tau$ and $\delta t$ are shown in units of $t_0$. Chain with 
$N=80$ sites, $(\alpha,\beta)=(1,0.15)$. 
See text for definitions of $N_{\rm sam}$, $t_0$, $\delta t$. 
}
\label{fig:cor_a1b15_nsvar}
\end{figure}

It must be noted that the inversion symmetry exhibited in 
Fig.~\ref{fig:cor_a1b15} reflects an ensemble average over
many samples, and is in general absent when one considers a single time 
interval,  as shown in the bottom panel of Fig.~\ref{fig:cor_a1b15_nsvar}.
The distinctive shape of Fig.~\ref{fig:cor_a1b15} only begins to
emerge upon accumulation of at least a few tens of samples, see 
top panel of Fig.~\ref{fig:cor_a1b15_nsvar}. 

\subsection{Domain wall approach}

We now investigate the extent to which an adaptation of domain-wall (DW) 
theory may be suitable for the present case. 
As is known~\cite{ksks98,ps99,ds00,sa02,rbsdq16}, the DW approach
incorporates fluctuations beyond the mean-field picture of TASEP. For 
one-dimensional systems with spatially uniform hopping and a single type of 
particle, it reproduces several exact results either exactly or to a very good 
numerical approximation. DW treatments have provided good quantitative account 
of non-stationary properties of single-species TASEP~\cite{sa02,rbsdq16}.

The flips between opposite symmetry-broken states exhibited by the systems
studied here constitute an extreme case of cooperative behavior, though
of course it originates in the usual ("microscopic") stochastic mechanisms of 
the standard TASEP. On the other hand, the DW description corresponds to a 
"macroscopic" view of the system state. For the single-species case this corresponds to
a narrow domain wall separating a domain on the left side,    
with uniform site occupation (local density) $\rho_L$ controlled by the injection 
rate, from another on the right with uniform site occupation $\rho_R$:
\begin{equation}   
\rho_L=\alpha\quad;\qquad\rho_R=1-\beta\ .
\label{eq:rhomrhop}
\end{equation}
The mean field currents in the two domains are respectively
\begin{equation}
J_L=\alpha(1-\alpha);\qquad J_R=\beta(1-\beta)\ . 
\label{eq:jmjp}
\end{equation}

However these steady state currents do not balance at the domain wall, if it
is stationary. This indicates the need to allow for stochastic motion of the domain wall.
One postulates that the TASEP process can be represented by the stochastic hopping of the
domain wall, with asymmetric hopping rates $D_L$, $D_R$ given by
\begin{eqnarray}
D_L=\frac{J_L}{\Delta \rho}=\frac{\alpha(1-\alpha)}{1-\alpha-\beta}\ ; \nonumber \\
D_R=\frac{J_R}{\Delta \rho}=\frac{\beta(1-\beta)}{1-\alpha-\beta}\ ,
\label{eq:dmdp}
\end{eqnarray}
where  $\Delta \rho \equiv \rho_R-\rho_L=1-\alpha-\beta$.

So the fluctuations present in DW theory arise from current- and density 
mismatches induced by boundary conditions~\cite{ksks98,ps99,ds00}.

In this context, one sees that the way to mimic a flip would be by starting the
evolution of two domain walls, one for each species, driven by boundary conditions
corresponding to a specific majority species, say $+$ particles,
and allowing the system to reach steady state; then, suddenly switch the boundary
conditions to those appropriate to the opposite majority species. The dynamics of the
approach to a new steady state under a sudden change in boundary conditions has
been investigated in Ref.~\onlinecite{sa02} for the single-species TASEP.     

The fact remains, however, that we will be switching from one fixed set of boundary
conditions to a different, but also fixed, one. 
It is not clear from the outset how this affects the description
of the flipping process, which is one that involves detailed local adjustment to 
constantly-changing, stochastically determined microscopic fluctuations.

In line with DW theory, one starts with the mean-field densities and currents for each
species, with adaptations for the coupling at the injection points. For 
completeness we reproduce results from Ref.~\onlinecite{efgm95} in 
Eqs.~(\ref{eq:jpm})--(\ref{eq:jrhopm}) below. With:
\begin{eqnarray}
J^+=\alpha(1-n_1^+-n_1^-)=\beta n_N^+ \nonumber \\
J^-=\alpha(1-n_N^+-n_N^-)=\beta n_1^- 
\label{eq:jpm}
\end{eqnarray}
and defining
\begin{eqnarray}
\alpha^+=\frac{\alpha(1-n_1^+-n_1^-)}{1-n_1^+}=\frac{J^+}{J^+/\alpha + J^-/\beta}\qquad
\nonumber \\
\alpha^-=\frac{\alpha(1-n_N^+-n_N^-)}{1-n_1^-}=\frac{J^-}{J^-/\alpha + J^+/\beta}\qquad
\nonumber \\
\beta^+=\beta^-=\beta \ ,\qquad\qquad\qquad\qquad
\label{eq:alphapm}
\end{eqnarray}
one gets two single-species processes within the mean field approximation.
Specializing to the hd/ld phase and assuming the $+$ particles to be in the majority, 
one finds
\begin{equation}
\alpha^- = \frac{1+\alpha}{2}-\frac{1}{2}\sqrt{(1+\alpha)^2-4\alpha\beta}\ .
\label{eq:alpham}
\end{equation} 
For this case the currents and respective densities in the bulk are:
\begin{eqnarray}
\  J^+=\beta(1-\beta)\ ;\quad\quad \rho^+=1-\beta\quad\quad
 \nonumber \\
J^-=\alpha^-(1-\alpha^-)\ ;\quad\quad \rho^- = \alpha^-\ .\quad\quad
\label{eq:jrhopm}
\end{eqnarray}
The corresponding expression for $\alpha^+$ can be found from Eq.~(\ref{eq:alphapm}).

So for the two single-species processes with the $+$ species in the majority, 
Eqs.~(\ref{eq:rhomrhop}),~(\ref{eq:jmjp}) translate into:
\begin{eqnarray}   
\rho_L^+=\alpha^+\quad;\qquad\rho_R^+=1-\beta\ ;\nonumber \\
J_L^+=\alpha^+(1-\alpha^+);\qquad J_R^+=\beta(1-\beta)\ ,
\label{eq:rholrhorp}
\end{eqnarray}
\begin{eqnarray}   
\rho_R^-=\alpha^-\quad;\qquad\rho_L^-=1-\beta\ ;\nonumber \\
J_R^-=\alpha^-(1-\alpha^-);\qquad J_L^-=\beta(1-\beta)\ .
\label{eq:rholrhorm}
\end{eqnarray}
The hopping rates $D_{L,R}^\pm$ follow from plugging Eqs.~(\ref{eq:rholrhorp}),~(\ref{eq:rholrhorm})
into Eqs.~(\ref{eq:dmdp}).

For the case with the $-$ species in the majority, it is $\alpha^+$ which is given by 
Eq.~(\ref{eq:alpham}); Eq.~(\ref{eq:jrhopm}) turns into
\begin{eqnarray}
\  J^-=\beta(1-\beta)\ ;\quad\quad \rho^-=1-\beta\quad\quad
 \nonumber \\
J^+=\alpha^+(1-\alpha^+)\ ;\quad\quad \rho^+ = \alpha^+\ .\quad\quad
\label{eq:jrhopmnew}
\end{eqnarray}
Similarly, the new $\alpha^-$ is found from Eq.~(\ref{eq:alphapm}) with the $J^+$, $J^-$ from
Eq.~(\ref{eq:jrhopmnew}).

We started the DW evolution with the $D_{L,R}^\pm$  corresponding to the $+$ species majority
and waited some time $t_{\rm sst}$ until the system reached a steady state. This can be
ascertained by checking whether the density profiles $\rho_\pm (n)$ (where $n=1, \dots  N$ 
denotes site position along the chain) remain stationary. The definition of "time" ($t^{\rm DW}$) 
in this case comes up in
that one solves the discrete-time version of the (continuous time)  differential equation
for the DW position. Then the elementary time interval $dt$ separates consecutive 
lattice-wide updates of the  probability distribution $P(x,t)$ of finding the DW at the bond between
sites $x=n$ and $x=n+1$ at time $t$. See Ref.~\onlinecite{rbsdq16} for a discussion of this point.
Here we used $dt=1$, which corresponds to equivalence between $t^{\rm DW}$
and the  elementary simulational time $t_0$ as defined in Sec.~\ref{sec:num-intro}.

\begin{figure}
{\centering 
\resizebox*{3.2in}{!}{\includegraphics*{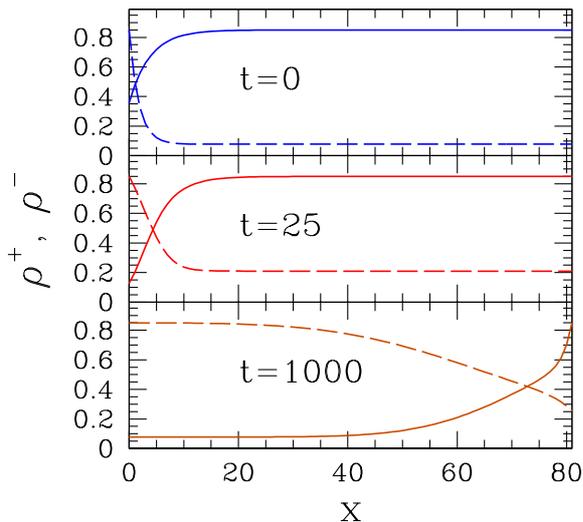}}}
\caption{
Density profiles for $+$ (full lines) and $-$ particles (dashed lines)
predicted by DW. At $t=0$ (upper panel) these are the steady-state ones for 
high density of $+$ particles. Then boundary conditions are instantaneously 
switched. Note rapid increase in bulk density of $-$ particles for very short times
(middle panel). By $t=1000$ (lower panel) the profiles are 
on the way to their new steady-state configurations. $(\alpha,\beta)=(1,0.15)$. 
Chain with $N=80$ sites at $x_n=n$, $n=1, \dots, N$ in lattice parameter units.
}
\label{fig:dw_a1b15sw}
\end{figure}

\begin{figure}
{\centering 
\resizebox*{3.2in}{!}{\includegraphics*{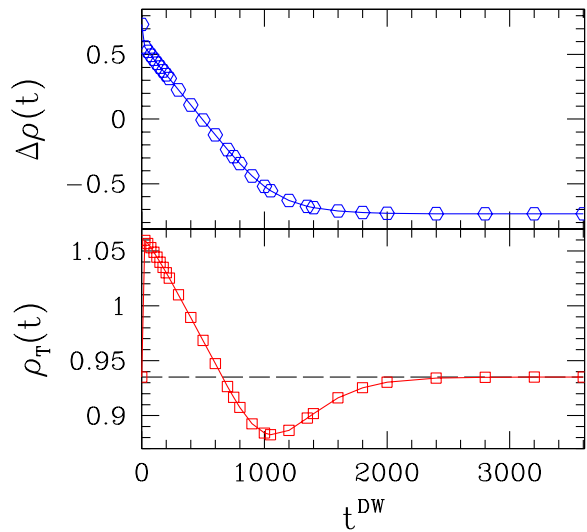}}}
\caption{
Density differences $\Delta\rho(t)$ (upper panel) and total densities $\rho_{\,T}(t)$ 
(lower panel) against time $t^{\rm DW}$ after boundary conditions are switched (see
Fig.~\ref{fig:dw_a1b15sw} and text) for chain with $N=80$ sites, $(\alpha,\beta)=(1,0.15)$.
The dashed line in the lower panel indicates $\rho_{\,T}(0)$. 
}
\label{fig:dw_dens}
\end{figure}

For $(\alpha,\beta)=(1,0.15)$ and $N=80$ sites we saw that $t_{\rm sst} \approx 10^4$ 
is enough to reach steady state to very good accuracy. We then reset the clock, 
and switched to boundary conditions such that the $-$ species majority became favored.
Fig.~\ref{fig:dw_a1b15sw} shows density profiles at specific times of interest.  
At $t=0$ when boundary conditions are switched (top panel) the profiles are the steady-state 
ones predicted by DW for $+$ particles in the majority. In the initial steps of evolution
according to  the new boundary conditions there is a very rapid inflow of $-$ particles:
their bulk density away from the left boundary goes from $\rho^-=0.07805$ (top panel)
to $\rho^-=0.20998$ (middle panel) by $t=25$. This is not accompanied by an equally dramatic 
emptying of the $+$ density, so initially the total density  $\rho_{\,T}$ actually grows.
Most of the change takes place in a step-function shape during the very first full-lattice update,
at the end of which the density variations have been $\delta\rho^+=-0.0063$, $\delta\rho^-=0.1310$.
The lower panel of the figure is for $t=1000$, which corresponds to the minimum value of 
$\rho_{\,T}$ during the flipping process, see Fig.~\ref{fig:dw_dens}.
By $t=3000$ (not shown) the profiles have become very close to their new steady-state configuration,
with deviations down to at most $0.05\%$ (this latter value corresponds to the new majority 
$-$ particles, near their injection end). 

Fig.~\ref{fig:dw_dens} depicts the evolution of $\Delta \rho$ and $\rho_{\,T}$,
showing that both these quantities become very close to their values 
associated with the new steady state by $t^{\rm DW} \approx 2500$; so
this can be broadly defined as the "duration" of the flip, as given by the
DW approach just described.

Recall that, from  the simulational results depicted in 
Fig.~\ref{fig:a1b15_trvar} one gets an estimate of $\approx 2\times 10^3$ in units of
$t_0$ for the duration of the flip exhibited there. Thus, accepting the identification
of $t^{\rm DW}$ with $t_0$ as argued above (see also Ref.~\onlinecite{rbsdq16}), 
this means that both in our implementation of DW and in numerical simulations
the flip duration is of the same order of magnitude, when measured in the
respective "time" units.

On the other hand, it is seen in the lower panel of Fig.~\ref{fig:dw_dens} that the minimum
value attained by the total density along the flipping process is $\rho_{\,T} \approx 0.88$,
only $6\%$ below its steady-state counterpart. This is a large discrepancy against
the simulational result shown in Fig.~\ref{fig:a1b15_trvar}, where $\rho_{\,T} \approx 0.2$
at the bottom of the dip is less than a quarter of the steady-state value. So the
"emptying" of the lattice, which is widely accepted as having a 
paramount role in the flipping mechanism, see 
Refs.~\onlinecite{efgm95,wsg05,gsw07} as well as the results
shown in Figs.~\ref{fig:a1b15_trvar} and ~\ref{fig:a1b15_trvar2} above, 
turns out to be a quantitatively minor feature in the DW result.    

Of course, as stated above, we are here mimicking a flip; the present result
indicates that the tools provided by DW theory do not fully emulate the large 
microscopic fluctuations driving a "real" flip. Incidentally it should be noted that
DW steady-state profiles, such as those on the top panel of Fig.~\ref{fig:dw_a1b15sw},
are in good agreement with both mean-field and simulational results~\cite{efgm95}. 

\section{Discussion and Conclusions} 
\label{sec:conc}

In this paper we studied the statistics and internal structure of the so-called
"flips"  exhibited by a specific implementation of a two-lane, two-species TASEP process in space 
dimensionality $d=1$, which follows the rules given in Secs.~\ref{intro} and~\ref{sec:tasep-theo}.
We focused on the region of the $(\alpha,\beta)$ parameter space where coexistence
between high-density and low-density (hd/ld) phases occurs. The above-mentioned flips consist 
of fluctuation-induced exchanges in composition of hd and ld phases (i.e.  between opposite 
broken-symmetry states of the two-species system). This  being a {\em bona fide} SSB phenomenon, 
the time between successive flips diverges in the thermodynamic limit. Thus we
considered only chains with a finite number $N$ of sites.

Initially we looked at time scales much longer than the typical duration of a flip.
We took the renormalized time scale $t_r=N_r\,t_0$ where $t_0$ is the elementary
time step corresponding to a full (stochastic) sequential  lattice update, with
$N_r=10^3$. On such scale the flips are essentially "instantaneous" for the values
of $(\alpha,\beta)$ and $N$ used. We collected the statistics of the times
$\Delta t_r$ elapsed between consecutive flips. For fixed $N=80$, $\alpha=1$,
$\beta$ in $[0.15,0.20]$ we found the distribution of the $\Delta t_r$ to be Poissonian
with $\beta$--dependent characteristic times $\tau_0 (\beta)$, 
indicating that flips are uncorrelated events on long time scales.
 
We showed in Fig.~\ref{fig:drhorms} how the evolution of the  $\tau_0 (\beta)$  is 
a good indicator of where in $(\alpha,\beta)$ parameter space a second-order 
transition is to be expected  (in this case, the transition hd/ld $\to$ ld~\cite{efgm95}).

Fourier analysis of our time sequences of $\Delta \rho (t)$ (see Fig.\ref{fig:fft}) 
shows clear evidence of some apparent periodic, low-frequency, structures in our simulational 
results for $\alpha=1$, $\beta=0.15$. 
For $\beta=0.20$ one can still distinguish a small maximum in the weight for 
larger-frequency components, again consistent with visual inspection of the raw data. 
At $\beta=0.225$ (closer to the transition at $\beta \approx 1/3$) our  
Fourier transform results show no clear trend.

In order to exploit the internal structure of flips, we then went to shorter
timescales, using $N_r=1$, $10$, and $10^2$,  see Figs.~\ref{fig:a1b15_trvar}
and~\ref{fig:a1b15_trvar2}. By doing so one detects relevant features which
may be lost in averages over scales for which flips are  instantaneous.  For
$N_r \leq 10^2$ one clearly sees the "emptying" of the lattice which is a qualitative 
feature broadly expected to be an integral part of the flipping 
process~\cite{efgm95,wsg05,gsw07}.   

%%%%%%%%%%%%%% 2nd referee, point (3) (1/2) %%%%%%%%%%%%%%%%%%%%%%%%%%%%%%%%
It is important to recall that the quantities exhibited in  
Figs.~\ref{fig:a1b15_trvar} and~\ref{fig:a1b15_trvar2} are always 
position-averaged particle densities (or density differences). 
Dealing with lattice-averaged quantities proved  invaluable in that it 
enabled us to work with smoothly-varying functions of "time".

We attempted to look at details of what happened  at individual sites, 
or even short subsections of the system; however, the sample-to-sample
fluctuations were so large as to prevent any clear conclusions
to be drawn.
%%%%%%%%%%%%%%%%%%%%%%%%%%%%%%%%%%%%%%%%%%%%%%%%%%%%%%%%%%%%%%%%%%%%%%%%%%%

We evaluated and analysed especially-designed two-time correlations which might be expected
to be sensitive to asymmetries between the time intervals before and after a flip takes
place. Such precursor events, if detected, would be qualitatively similar to the
mechanism described in Refs.~\onlinecite{wsg05} and ~\onlinecite{gsw07}.
In that case, it was shown that an initially empty latice (or any other initially symmetric 
configuration) is led into a symmetry-broken state through an "amplification loop" of initial 
fluctuations. An important difference is that in our case we studied flips between 
broken-symmetry states. We found strong indications that pre- and post-flip fluctuations
are essentially equivalent, see Fig.~\ref{fig:cor_a1b15}; no sign was found of precursor 
events to flips.

%%%%%%%%%%%%%% 2nd referee, point (3) (2/2) %%%%%%%%%%%%%%%%%%%%%%%%%%%%%%%%
On a complementary note to the discussion of lattice-averaged densities
two paragraphs above, we were able to get reasonably
smooth data even for a "single" sample of $C(\tau)$, see the bottom panel 
of Fig.~\ref{fig:cor_a1b15_nsvar}, because the quantity under
observation depended only on the evolution of the position-averaged density
(although it involved a short time interval).
%%%%%%%%%%%%%%%%%%%%%%%%%%%%%%%%%%%%%%%%%%%%%%%%%%%%%%%%%%%%%%%%%%%%%%%%%%%

We also used a domain wall (DW) approach to mimic a flip, by suddenly reversing the boundary 
conditions applied to a system in an SSB steady state.
One notable feature emerging from the DW evolution 
is that the overall duration of the flipping process  is similar, when measured in the
respective time units ($t^{\rm DW}$), to that of a simulationally generated flip (measured
in computational time $t_0$).

On the other hand, the DW evolution fails to provide a quantitatively adequate account 
of the lattice emptying process which has been seen in simulations, and which is 
theoretically understood to be a fundamental component of the flipping process. 
Numerically, in the DW description the total density goes through a minimum only a few 
percentage points below its steady-state value, while simulations would lead one to
expect a typical maximum dip of order $70-80\%$ (for the same $\alpha$, $\beta$, and $N$).
See Fig.~\ref{fig:dw_dens}.

\begin{acknowledgments}
S.L.A.d.Q. thanks the Rudolf Peierls Centre for Theoretical Physics, 
Oxford, for hospitality during his visit.
The research of S.L.A.d.Q. is supported by the Brazilian agencies
Conselho Nacional de Desenvolvimento Cient\'\i fico e
Tecnol\'ogico  (Grant No. 303891/2013-0)
and Funda\c c\~ao de Amparo \`a Pesquisa do Estado do Rio
de Janeiro (Grants Nos. E-26/102.760/2012, E-26/110.734/2012, and 
E-26/102.348/2013).
\end{acknowledgments}

\end{document}